\documentclass[letterpaper]{JHEP3}
\usepackage{epsfig}
\usepackage{graphicx}
\usepackage{subfigure}
\usepackage{dcolumn}
\usepackage{bm}
\usepackage{float}
\usepackage{cancel}

\def\beq{\begin{equation}}
\def\eeq{\end{equation}}
\def\bea{\begin{eqnarray}}
\def\eea{\end{eqnarray}}
\def\to{\rightarrow}
\def\go{\tilde{g}}

\preprint{IPMU10-0197}

\title{
\vspace*{0.75cm} {NLSP Gluino Search at the Tevatron and early LHC}
}
\author{M. Adeel Ajaib$^1$\footnote{adeel@udel.edu}, Tong Li$^1$\footnote{tli@udel.edu}, Qaisar Shafi$^1$\footnote{shafi@bartol.udel.edu}, Kai Wang$^2$\footnote{kai.wang@ipmu.jp}\\
{\it $^1$Bartol Research Institute, Department of Physics and Astronomy,
University of Delaware, Newark, DE 19716, USA} \\
{\it $^2$IPMU, University of Tokyo, Kashiwa, Chiba 277-8568, JAPAN}}

\date{\today}

\abstract{We investigate the collider phenomenology of gluino-bino
co-annihilation scenario both at the Tevatron and 7 TeV LHC. This
scenario can be realized, for example, in a class of realistic
supersymmetric models with non-universal gaugino masses and
$t-b-\tau$ Yukawa unification. The NLSP gluino and LSP bino should
be nearly degenerate in mass, so that the typical gluino search
channels involving leptons or hard jets are not available.
Consequently, the gluino can be lighter than various bounds on its
mass from direct searches. We propose a new search for NLSP gluino
involving multi-b final states, arising from the three-body decay
$\tilde{g}\to b\bar{b}\tilde{\chi}_1^0$. We identify two realistic
models with gluino mass of around 300 GeV for which the three-body
decay is dominant, and show that a 4.5 $\sigma$ observation
sensitivity can be achieved at the Tevatron with an integrated
luminosity of 10~fb$^{-1}$. For the 7 TeV LHC with 50~pb$^{-1}$ of
integrated luminosity, the number of signal events for the two
models is $\cal O$(10), to be compared with negligible SM
background events.}

\keywords{NLSP gluino, hadronic colliders}

\begin{document}

\section{Introduction}

Low-scale supersymmetry (SUSY) is arguably the leading candidate of
new physics and will be seriously tested at the CERN Large Hadron
Collider (LHC). It provides an elegant solution to the gauge
hierarchy problem and with conserved $R$-parity, the thermal relic
abundance of the lightest neutralino (LSP) can often be identified
with dark matter, consistent with the current cosmological
observations~\cite{wmap}. The recent results reported by the CDMS-II
experiment may indicate the presence of dark matter with mass of
around ${\cal O}(100)~{\rm GeV}$~\cite{cdms}. To reproduce the
required thermal relic abundance, a pure wino or Higgsino dark
matter should have mass of a few TeV due to rapid annihilation rate.
On the contrary, the annihilation rate of a pure bino with mass of
around 100~GeV is too slow, and leads to excessive relic
abundance~\cite{Yaguna}. A variety of constraints from low energy
flavor physics and CP violating physics typically favor scenarios
with heavy sfermions~\cite{cohen,feng,arkani}. Consequently, pure
bino self annihilation is further suppressed if scalars in the
$t$-channel become heavier. To enhance the annihilation rate of pure
bino, there are generally two categories of models available, and
both will lead to interesting phenomenology at the LHC. In one
scenario the dark matter is a bino-wino or bino-Higgsino
mixture~\cite{well-temper}, so that
$\tilde{\chi}^{0}_{1}\tilde{\chi}^{0}_{1}$ annihilation is enhanced
via the enlarged coupling. The mass degeneracy $M_{1}\simeq M_{2}$
leads to a nearly degenerate chargino-neutralino
spectrum~\cite{Giudice:2010wb}. The second scenario is
co-annihilation~\cite{coann,profumo}, and if the scalars are heavy, the
required relic abundance can be achieved through co-annihilation
with a strongly-interacting particle such as the gluino, namely
$\tilde{\chi}^{0}_{1}\go,\go\go\to
f\bar{f}$~\cite{Yaguna,Gogoladze:2009bn,Ilia,zuowei}. This scenario
can be realized in a realistic class of $t-b-\tau$ Yukawa unified
models~\cite{Gogoladze:2009bn,Ilia,Anant}. In the gluino-bino
co-annihilation scenarios, the gluino is the
next-to-lightest-supersymmetric-particle (NLSP), and the mass
splitting betwen $\tilde{g}$ and $\tilde{\chi}_1^0$ is relatively
small, namely~\cite{Yaguna} \beq
\frac{M_{\go}-M_{\tilde{\chi}_{1}^0}}{M_{\tilde{\chi}_1^0}} \lesssim
20\%~. \label{co} \eeq
Note that the Sommerfeld enhancement of bino-gluino co-annihilation cross section is rather mild, such that Eq.~(\ref{co}) is altered by only $2-3\%$~\cite{profumo,zuowei}.

The relations between gaugino masses will be crucial in understanding the nature of supersymmetry breaking and of the underlying theory
at ultra-high energy scale. For instance, supersymmetric grand unification, string dilaton SUSY breaking,
and minimal gauge mediation all predict the gaugino mass relations
$M_3/{g_3^2} = M_2/{g_2^2} = M_1/{g_1^2}$.
With NLSP gluino, one must invoke non-universal gaugino masses at $M_{GUT}$.
This not only implies very different physics from
the above models but can also give rise to very different phenomenology at hadron colliders. We focus in this paper
on the phenomenological implications of bino-gluino co-annihilation scenarios at hadron colliders.

Being a color octet fermion, gluino pair production is the most promising discovery
channel for supersymmetry at hadron colliders. In an environment such as this with huge QCD jet backgrounds,
isolated charged leptons ($\mu^{\pm},~e^{\pm}$) and $b$-jets usually play an important role in the searches.
For the most well studied scenarios where the charginos are lighter than gluinos, the Majorana nature of the gluino will result in
same-sign chargino signature (jets plus $\tilde{\chi}^{\pm}\tilde{\chi}^{\pm}$).
This eventually leads to same-sign dileptons plus jets with very little Standard Model (SM) background.
With a NLSP gluino, however, the chargino as well as the leptons are absent in the gluino cascade decay. The conventional search strategy does not work here and the NLSP gluino can evade the current bounds from direct searches at the Tevatron. The NLSP gluino can be relatively light and its production rates can be large both at the Tevatron and
especially at the LHC.

On the other hand, as a consequence of $t-b-\tau$ Yukawa unification,
the third generation squarks, stops or sbottoms, are usually much
lighter than those of the first two generations.
In the large $\tan\beta$ limit, sbottoms are often the lightest squarks.
As a consequence, gluino decays may lead to top-rich or bottom-rich events.
In the co-annihilation region with $\Delta M = M_{\go}-M_{\tilde{\chi}^0_{1}} \ll 2 m_{t}$,
there is no phase-space for on-shell top quarks. The gluino decay into
$b$-jets, $\go \to b\bar{b}\tilde{\chi}^{0}_{1}$, then becomes dominant
and this enable us to search for NLSP gluino via multi-$b$ jets, namely
\beq
p\bar{p},pp\to \go \go \to b\bar{b}b\bar{b} +\cancel{E}_{T}.
\eeq
Multi-$b$ events have been widely proposed for light scalar searches,
for instance in NMSSM with the Higgs decaying into multi-$b$ jets via light scalars ($h\to aa\to b\bar{b}b\bar{b}$)~\cite{han}.
The multi-$b$ jets plus significant missing transverse energy may only appear in $Wh,Zh$ associated
productions whose production rates are much smaller than gluino pair production. The reconstruction
is also very different.

The paper is organized as follows.
In the next section (II), we discuss NLSP gluino decay
and the parameter space where multi-$b$ jet production is significant. We very
briefly comment models in which NLSP gluino can be realized
and discuss its implications. In Section III and IV, we study the collider phenomenology of this scenario, with event selection and identification both at Tevatron and the LHC. We summarize our findings in Section V.

\section{NLSP Gluino: Decays and Benchmark Models}

As mentioned earlier, the gluino-bino co-annihilation scenario
requires the gluino to be NLSP in the sparticle spectrum, and to be
nearly degenerate in mass with the bino LSP. The mass difference
between the two should be $\lesssim 20\% M_{\tilde{\chi}^0_{1}}$. In
the framework of minimal supergravity, this clearly requires
non-universal gaugino masses at $M_{GUT}$. The leading motivations
for grand unification theories (GUTs such as SO(10)) are the
explanation of tiny neutrino masses and charge quantization.
However, a partial unified model such as $SU(4)_{C}\times
SU(2)_{L}\times SU(2)_{R}$ (4-2-2)~\cite{Pati:1974yy} also provides
solutions to both problems. Non-universal asymptotic gaugino masses
are naturally accomodated in the supersymmetric 4-2-2 model and have
recently been investigated in Ref.~\cite{Gogoladze:2009bn,Ilia}.
Other examples with non-universal gaugino masses include a
supersymmetric $SU(5)\times SU(3)_{\rm Hypercolor}$ proposed to
explain doublet-triplet splitting problem in $SU(5)$
GUT~\cite{moroi,doublet-triplet}, and GUT models with non-singlet
$F$-term vevs~\cite{Martin}.

Since we focus on a spectrum with NLSP gluino, on-shell charginos are
kinematically forbidden in gluino decay.
The color octet gluino therefore can decay only into colored SM particles such as the gluon octet or a $q\bar{q}$ pair, plus the color singlet LSP $\tilde{\chi}^0_{1}$,
\beq
\go \to q \bar{q} \tilde{\chi}^0_{1}, g \tilde{\chi}^0_{1}.
\eeq
The three-body decay $\go\to q\bar{q} \tilde{\chi}^0_{1}$ is through off-shell squark interchange, while the two-body decay
$g\tilde{\chi}^0_{1}$ can be realized by the triangle loop involving squarks.
The partial widths of these two decay channels are given by~\cite{loop1,loop}
\begin{eqnarray}
\Gamma(\tilde{g}\to g \tilde{\chi}^0_{1})&=&
{(M_{\tilde{g}}^2-M_{\tilde{\chi}_1^0}^2)^3\over 2\pi M_{\tilde{g}}^3}[{g_3^2g_1\over 128\pi^2}(M_{\tilde{g}}-M_{\tilde{B}})\sum_qQ_q({1\over M_{\tilde{q}_L}^2}-{1\over M_{\tilde{q}_R}^2})N_{1B}\nonumber \\
&+&{g_3^2y_t^2\over 32\sqrt{2}\pi^2\sin\beta}
({1\over M_{\tilde{q}_L}^2}+{1\over M_{\tilde{u}_R}^2})N_{1H_u}v(1+{\rm ln}{m_t^2\over M_{\tilde{g}}^2})]^2,
\label{loopdecay}\\
\Gamma(\tilde{g}\to q\bar{q} \tilde{\chi}^0_{1})&=&
{M_{\tilde{g}}^5\over 768\pi^3}[({g_3g_1\over 6 M_{\tilde{q}_L}^2}N_{1B}+{g_3g_2\over 2 M_{\tilde{q}_L}^2}N_{1W})^2+({2g_3g_1\over 3M_{\tilde{u}_R}^2}N_{1B})^2\nonumber \\
&+&({g_3g_1\over 6 M_{\tilde{q}_L}^2}N_{1B}-{g_3g_1\over 2 M_{\tilde{q}_L}^2}N_{1W})^2+({g_3g_1\over 3M_{\tilde{d}_R}^2}N_{1B})^2]f({M_{\tilde{\chi}^0_{1}}\over M_{\tilde{g}}}) \ (q=u,d),\\
f(x)&=&1+2x-8x^2+18x^3-18x^5+8x^6-2x^7-x^8\nonumber \\
&-&12x^4{\rm ln}x^2+12x^3(1+x^2){\rm ln}x^2.
\label{treedecay}
\end{eqnarray}
Here $N_{1B}$, $N_{1W}$ and $N_{1H_u}$ are respectively the bino $\tilde{B}$, wino $\tilde{W}$ and Higgsino $\tilde{H}_u$ components of the LSP neutralino $\tilde{\chi}_1^0$.
The three-body decays will be suppressed if the scalar masses are too large, or due to phase space if the mass difference between $\tilde{g}$ and $\tilde{\chi}_1^0$ is too small. The two-body mode may dominate in this case. If either the gluino two-body decay $\go\to g\tilde{\chi}^0_{1}$, or three-body decay due to the small mass difference into light jets dominates, the final state jets are typically as soft as
those from parton showers. In this case the gluino decay very likely gets buried in the
huge QCD background, and consequently the search becomes extremely challenging~\cite{zuowei, Alwall:2008ve}.
Our study will focus on a different region where the gluino three-body decays
into multi-$b$ jets dominate.

\begin{figure}[tb]
\centering
\includegraphics{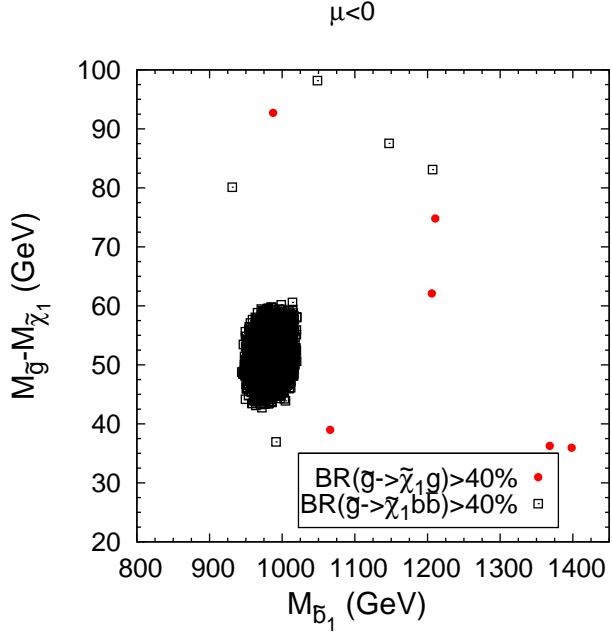}
\caption{Mass difference $M_{\tilde g}-M_{\tilde{\chi}_1^0}$ versus
$M_{\tilde{b}_1}$ for 4-2-2 model with $\mu<0$. The points shown
satisfy all the experimental constraints described in the text. The
red circular points depict scenarios for which the branching
fraction of two-body decay is dominant, i.e. ${\rm BR}({\tilde g}
\to g\tilde{\chi}_1^0)>40\%$. For the empty square points in black
the three-body decay is dominant, ${\rm BR}({\tilde g} \to b \bar{b}
\tilde{\chi}_1^0)>40\%$. Gluino is the NLSP for all the points shown
in the figure.} \label{decay}
\end{figure}

To illustrate the search strategy, we have selected two benchmark
points from a previous study of supersymmetric 4-2-2
models~\cite{Gogoladze:2009bn,Ilia}. In 4-2-2 models, the matter
fields of each family belong in $(4,2,1)$ and $(4,1,2)$. The third
family fermion masses, to a good approximation, arise from a Yukawa
coupling to the bi-doublet $(1,2,2)$. Thus, $t-b-\tau$ Yukawa
unification arises as a natural
prediction~\cite{Gogoladze:2009bn,Ilia,Anant} \beq
Y_{t}=Y_{b}=Y_{\tau}\equiv Y_{\rm Dirac}~. \eeq In SO(10),
$t-b-\tau$ Yukawa unification typically predicts gluino to be the
lightest colored sparticle~\cite{BaerYukawa}, while in 4-2-2 models
one realizes the NLSP gluino scenario through gluino-bino
co-annihilation. A large bottom Yukawa $Y_{b}$ also naturally drives
sbottom to be the lightest squark. With $\Delta M\simeq 50$~GeV and
$\cal O$(TeV) squarks, $\go\to b\bar{b}\tilde{\chi}^0_{1}$ decay
often dominates. Figure~\ref{decay} shows the dependence of the
gluino decay BR in the $\Delta M-M_{\tilde{b}_{1}}$ plane for the
4-2-2 model with $\mu<0$. We require consistency of the model with
various phenomenological constraints such as ${\rm BR}(B_s \to \mu^+
\mu^-) $, ${\rm BR}(b\to s \gamma)$, ${\rm BR}(B\to \tau \nu)$,
$\Delta(g-2)_{\mu}$, WMAP relic density (in the 5$\sigma$ range),
and all the sparticle mass bounds~\cite{Gogoladze:2009bn,Ilia}. In
Fig.~\ref{decay} the points which satisfy $t-b-\tau$ Yukawa
unification are a subset of the displayed points and mostly lie in
the dense region around 1~TeV sbottom mass, where the mass
difference $\Delta M$ is between 40 and 60 GeV. For this region the
branching fraction of gluino three-body decay ${\rm BR}(\go\to
b\bar{b}\tilde{\chi}^0_{1})$ dominates both the two-body one, ${\rm
BR}(\go\to g\tilde{\chi}^0_{1})$, as well as the three-body decay
into light quarks ${\rm BR}(\go\to q\bar{q}\tilde{\chi}^0_{1})$. For
a sufficiently large sbottom mass the two-body decay is dominant, as
can be seen from Fig.~\ref{decay}. Note that we show only those
scenarios in Fig.~\ref{decay} for which gluino is the NLSP. We have
picked two benchmark models, one with $\mu>0$ (Model A), listed as
point 1 in Ref.~\cite{Gogoladze:2009bn}; the second model has
$\mu<0$ (Model B), selected from the large number of points in
Fig.~\ref{decay}. The relevant observables are listed in
Table~\ref{model}. Both models can evade the direct search Tevatron
gluino bounds due to the dominant decay mode $\tilde{g}\to b \bar{b}
\tilde{\chi}_1^0$.

\begin{table}[tb]
\begin{center}
\begin{tabular}{|c|c|c|}
\hline
& Model A ($\mu>0$) & Model B ($\mu<0$) \\
\hline
$m_0$ (GeV) & 14110 & 1513\\
\hline
$M_1$ (GeV) & 499.54 & -479.49\\
\hline
$M_2$ (GeV) & 832.03 & -845.5\\
\hline
$M_3$ (GeV) & 0.7945 & 69.53\\
\hline
$\tan\beta$ & 50.82 & 47.7\\
\hline
$A_0$ & -34551.2 & -1668.84\\
\hline
$m_{H_u}$ (GeV) & 6092.74 & 492.41\\
\hline
$m_{H_d}$ (GeV) & 14194.5 & 1071.75\\
\hline
\hline
$M_{\go}$ (GeV) & 329 & 261\\
\hline
$M_{\tilde{\chi}^0_{1}}$ (GeV) & 284 & 207\\
\hline
$M_{\tilde{b}_{1}}$ (GeV) & 5294 & 950\\
\hline
BR($\go \to b\bar{b} \tilde{\chi}^0_{1}$) & 76.3\% & 50.8\%\\
\hline
\end{tabular}
\end{center}
\caption{Model parameters at GUT scale (above double line) and low scale (below double line) for two benchmark models. Note that the bino component of $\tilde{\chi}_1^0\geq 99.9\%$.}
\label{model}
\end{table}

\section{NLSP gluino Search at the Tevatron}
In both benchmark models, the gluino masses are of order 200-300 GeV, and
their pair production rates at the Tevatron are around the pico-barn level.
Therefore, a search for a relatively light NLSP gluino appears quite promising. In this section, we illustrate
how one could identify NLSP gluinos at the Tevatron for the two benchmark models above.

As shown in Table~\ref{model} for both models we focus on NLSP gluino with the dominant three-body decay
$\tilde{g}\to b\bar{b}\tilde{\chi}_1^0$.
Therefore we wish to
identify signal events of gluino pair production with 4 $b$-jets plus missing transverse
energy $\cancel{E}_{T}$,
\begin{eqnarray}
p\bar{p}\to \tilde{g}\tilde{g}\to b\bar{b}b\bar{b}+\cancel{E}_T.
\end{eqnarray}
Due to the relatively long lifetime of the $B$-mesons,
their decays on average take place $\cal O$(mm) distance away from
the primary interacting vertex. With the vertex detector, tagging
jet with decaying $B$-mesons will significantly reduce the
QCD jets background. The $b$-tagging efficiency at the Tevatron is taken
to be $\epsilon_{b} = 50\%$~\cite{cdf}.
The $b$-jet production in the SM is either due to gluon splitting or from top quark
decay originating from top pair and/or single production ($t\bar{t}$ pair provides $b\bar{b}$ in the final states, while single top production
$j g\to \bar{b} t j^{\prime}$ also provides $b\bar{b}$), and so
the $b$-jets always arise as pairs. Hence we only need to tag three $b$-jets
so that we do not have to pay the additional 50\% loss in the fourth $b$-tagging.
After multiplying by the $b$-tagging efficiency,
both the signal and the background events are reduced by
\beq
\epsilon^{3}_{b} = 12.5\%~.
\eeq
By requiring 3 $b$-tagged jets, the SM production $4b+X$ becomes the leading
irreducible background. Also, there exist reducible backgrounds due to
other jets being mis-identified as $b$-events. About 15\% of $D$-mesons
in the jets can be mis-identified as $B$'s, and the mis-$b$ tagging rate for light jets
is 0.4\%~\cite{cdf}. Since the $c$-jet production in the SM is very similar to the $b$-jet production, the
production rate is at comparable level. With the 15\% faking rate, we can safely neglect
the $c$-jet production in our study. However, the production rate of light jets is several orders of
magnitude higher than that of pure $b$-jet production~\cite{MGME} and cannot therefore be neglected even with 0.4\% mis-$b$-tagging rate.
Therefore, we include $jjb\bar{b}+X$ in our background analyses.

Besides the $b\bar{b}$, the dark matter particles $\tilde{\chi}^0_{1}$ also appear in the gluino cascade decay.
The missing transverse energy $\cancel{E}_{T}$ is another characteristic feature of the signal.
The irreducible SM background for $\cancel{E}_{T}$ is from $Z$ production, with the branching fraction of $Z$ invisible decay ($Z\to \nu\bar{\nu}$) as 20\%.
However, due to
the uncertainty of mis-measurement in jet energy or momentum, the events without
$Z$ can also induce $\cancel{E}_{T}$. The third source is due to leptonic decays of $W^{\pm}$ bosons, especially $W^{\pm}\to\tau^{\pm}\nu_{\tau} \to \ell^{\pm} \nu_{\ell}\bar{\nu}_{\tau}\nu_{\tau}$ where
the leptons from $\tau$ three-body decays are below the visible lepton cut ($p^{\ell}_{T}>10~$GeV).
The SM backgrounds that we consider in the study are then
\begin{eqnarray}
& b\bar{b}b\bar{b}, jjb\bar{b},\nonumber\\
& b\bar{b}b\bar{b}Z, jjb\bar{b}Z ~~~{\rm with}~~{\rm BR}(Z\to \nu\bar{\nu})=20\%,\nonumber\\
& t\bar{t}\to b\bar{b}jj \tau^{\pm}\nu_{\tau}~~~~{\rm
with}~~\tau^{\pm}~~{\rm leptonic \ decay}~.
\end{eqnarray}
To simulate the detector effects, we smear the hadronic jet energy
by a Gaussian distribution whose width is parameterized as
\cite{cdf} \beq \frac{\Delta E_{j}}{E_{j}}=
\frac{75\%}{\sqrt{E_{j}/\rm GeV}}\oplus 5\%. \label{cdfsmear} \eeq
The following basic kinematical cuts on the transverse momentum
($p_T$), the pseudo-rapidity ($\eta$), and the separation in the
azimuthal angle-pseudo rapidity plane ($\Delta R=\sqrt{(\Delta
\phi)^2+(\Delta \eta)^2}$) between two jets have been employed for
jet-selection ~\cite{cdf} \beq p^{j}_{T} > 15~{\rm GeV}, |\eta_{j}|<
1.0~, \Delta R_{jj}>0.4. \label{basic} \eeq

\begin{figure}[tb]
\centering
\includegraphics[width=0.45\textwidth]{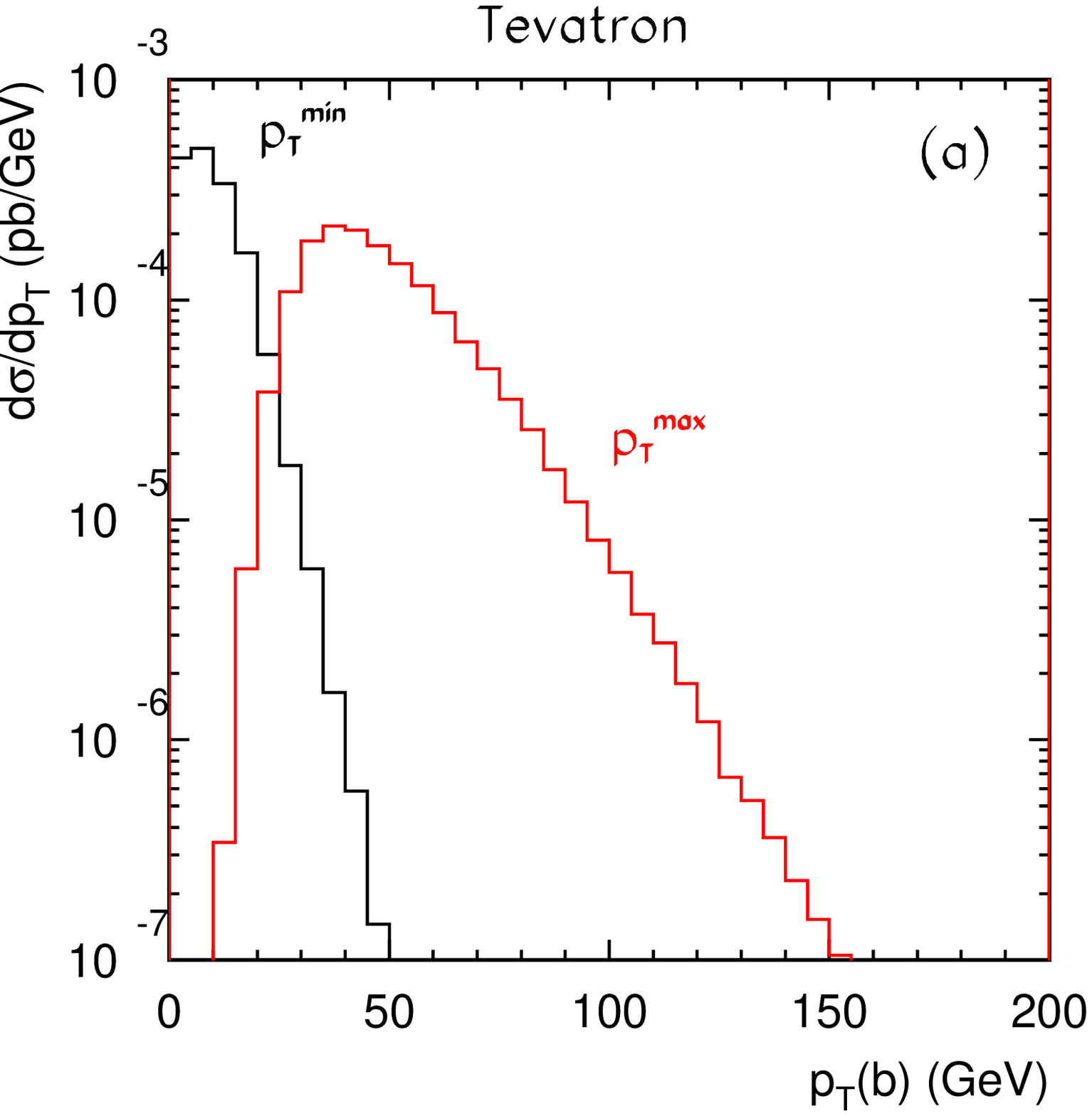}
\includegraphics[width=0.45\textwidth]{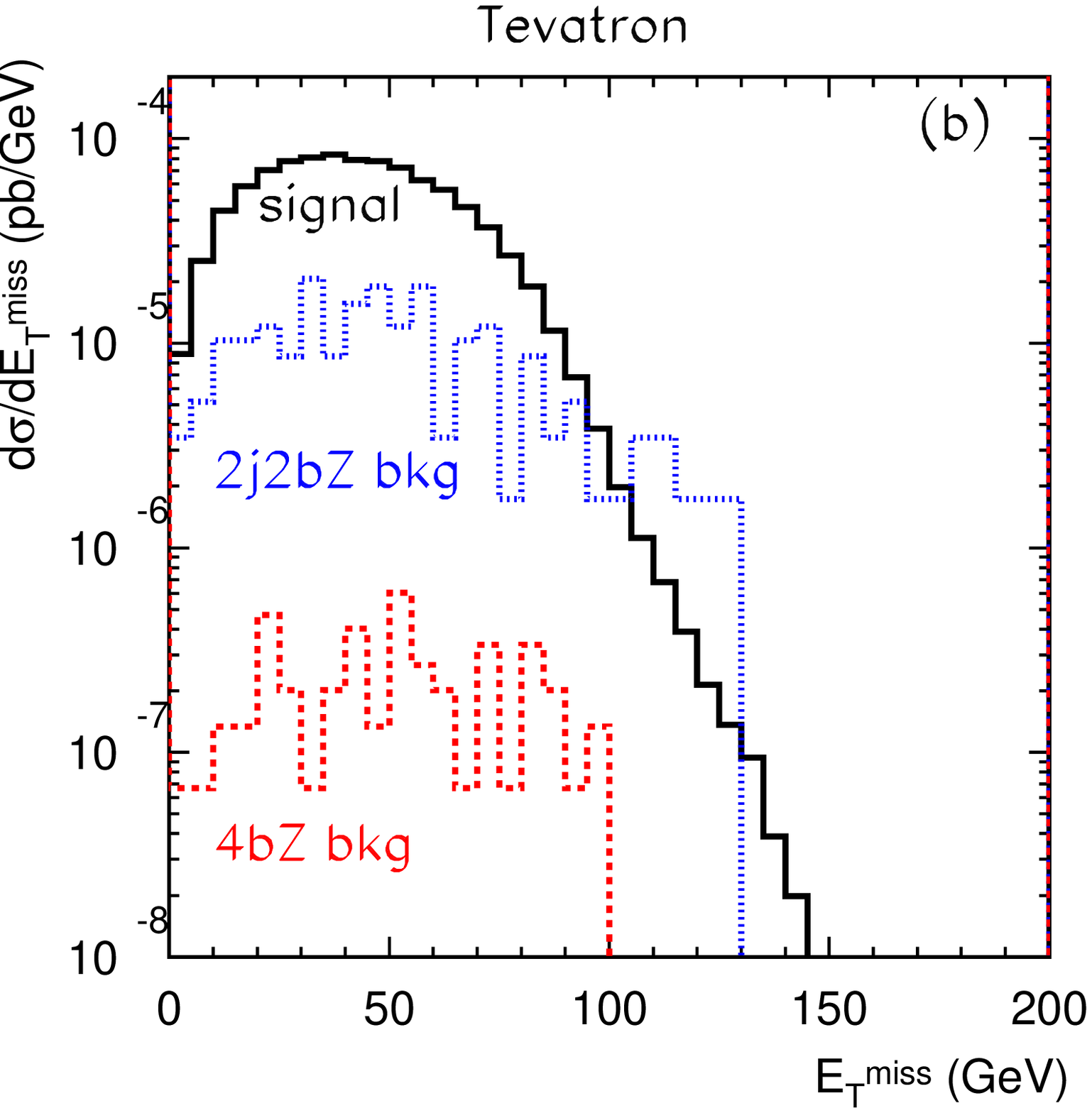}
\caption{(a) Minimal and maximal $p_{T}$ distribution of 4 $b$-jets
in the signal $p\bar{p}\to \go\go\to
b\bar{b}b\bar{b}\tilde{\chi}^{0}_{1}\tilde{\chi}^{0}_{1}$ at
Tevatron. (b) Missing transverse energy ($\cancel{E}_{T}$)
distribution in signal events as well as its background $4bZ$ and
$jjb\bar{b}Z$ at Tevatron, taking the branching fraction of $Z$
invisible decay ($Z\to\nu\bar{\nu}$) as 20\% and mis-$b$-tagging
rate of light jet as 0.4\%. } \label{missinget}
\end{figure}

Because of the relatively small mass difference $\Delta
M=M_{\go}-M_{\tilde{\chi}^0_{1}}\simeq 50$~GeV, the $b$-jets in
final states could be rather soft. The minimal and maximal $p_T$
distributions of $b$-jets are plotted in Fig.~\ref{missinget}(a).
The plots that we show in this paper are all for benchmark model B,
with the relevant features for Model A expected to be very similar.
One can see that the $b$-jet with minimal $p_T$ would be rejected by
the basic $p_T$ cut in Eq.~(\ref{basic}). To retain as many signal
events as possible, we apply the $p_T$ cut for three $b$-jets except
the softest one. The softest jet will be viewed as unreconstructed
calorimeter energy in the detector. It is consistent with the
requirement of 3 tagged $b$-jets above. The missing transverse
energy $\cancel{E}_T$ is reconstructed according to the smeared
observed particles, namely $b$-jets in our case. We show the
$\cancel{E}_{T}$ distribution of the signal and backgrounds
including basic cuts for $\eta$ and $\Delta R$ in Eq.~(\ref{basic})
and $b$-jet and mis-$b$ jet tagging efficiency in
Fig.~\ref{missinget}(b). We find that by requiring a significant
$\cancel{E}_{T}$ cut in the final states, namely \beq \cancel{E}_{T}
> 30~{\rm GeV}, \eeq the SM $4b$ and $jjb\bar{b}$ backgrounds at
Tevatron can be completely eliminated. Since the signal events do
not contain any lepton in the final states, we veto any event with
visible leptons of \beq p^{\ell}_{T} > 10~{\rm GeV}~. \eeq The
leading reducible background of soft leptons type is due to
semi-leptonic $t\bar{t}$ events with one $W^{\pm}$ decaying into
$\tau^{\pm}$, and $\tau^{\pm}$ further decaying into soft leptons.
With one of the light jets faking a $b$-jet, we find that the
contribution due to $t\bar{t}$ is below 0.01 fb. The leading
irreducible background after $\cancel{E}_{T}$ cut is then
$jjb\bar{b}Z$ with invisible $Z$ decay.

\begin{table}[tb]
\begin{center}
\begin{tabular}[t]{|c|c|c|c|c|c|c|}
  \hline
  $\sigma(\rm fb) @$ Tevatron & Model A & Model B & $4b$ & $4bZ$ & $2j2bZ$ & $S/\sqrt{B}$\\
  \hline
  basic cuts and 3b tagging & 2.3 & 4.8 & $2.7\times 10^3$ & 0.02  & 1 &\\
  \hline
  $\cancel{E}_T>30$~GeV & 1.4 & 3.3 & $-$ & 0.019 & 0.95 & 4.5(A)/11(B) \\
  \hline
\end{tabular}
\end{center}
\caption{Production cross section for $p\bar{p}\to
\tilde{g}\tilde{g}\to b\bar{b}b\bar{b}
\tilde{\chi}_1^0\tilde{\chi}_1^0$ in models A and B and for
backgrounds $4b,4bZ,2j2bZ$, after basic cuts and missing energy cut
at the Tevatron with a luminosity of 10~fb$^{-1}$. }
\label{tevatron}
\end{table}

In Table~\ref{tevatron}, we summarize the signal and background
cross sections at the Tevatron for the two benchmark models, after implementing the basic cuts and $\cancel{E}_T$ cut.
For our numerical analyses, we use the CTEQ6L1 parton distribution
function~\cite{pdf}. The SM backgrounds are simulated by the automatic
package Madgraph/Madevent~\cite{MGME}.
The signal significance is obtained in terms of Gaussian statistics, given by the ratio $S/\sqrt{B}$ of signal and background events with a luminosity of 10~fb$^{-1}$. For benchmark model A with $M_{\tilde{g}}\sim 300$ GeV, the statistical significance is close to 5$~\sigma$, while for benchmark model B with a smaller gluino mass, as expected it is 2.5 times larger.

\begin{figure}[tb]
\centering
\includegraphics[width=0.45\textwidth]{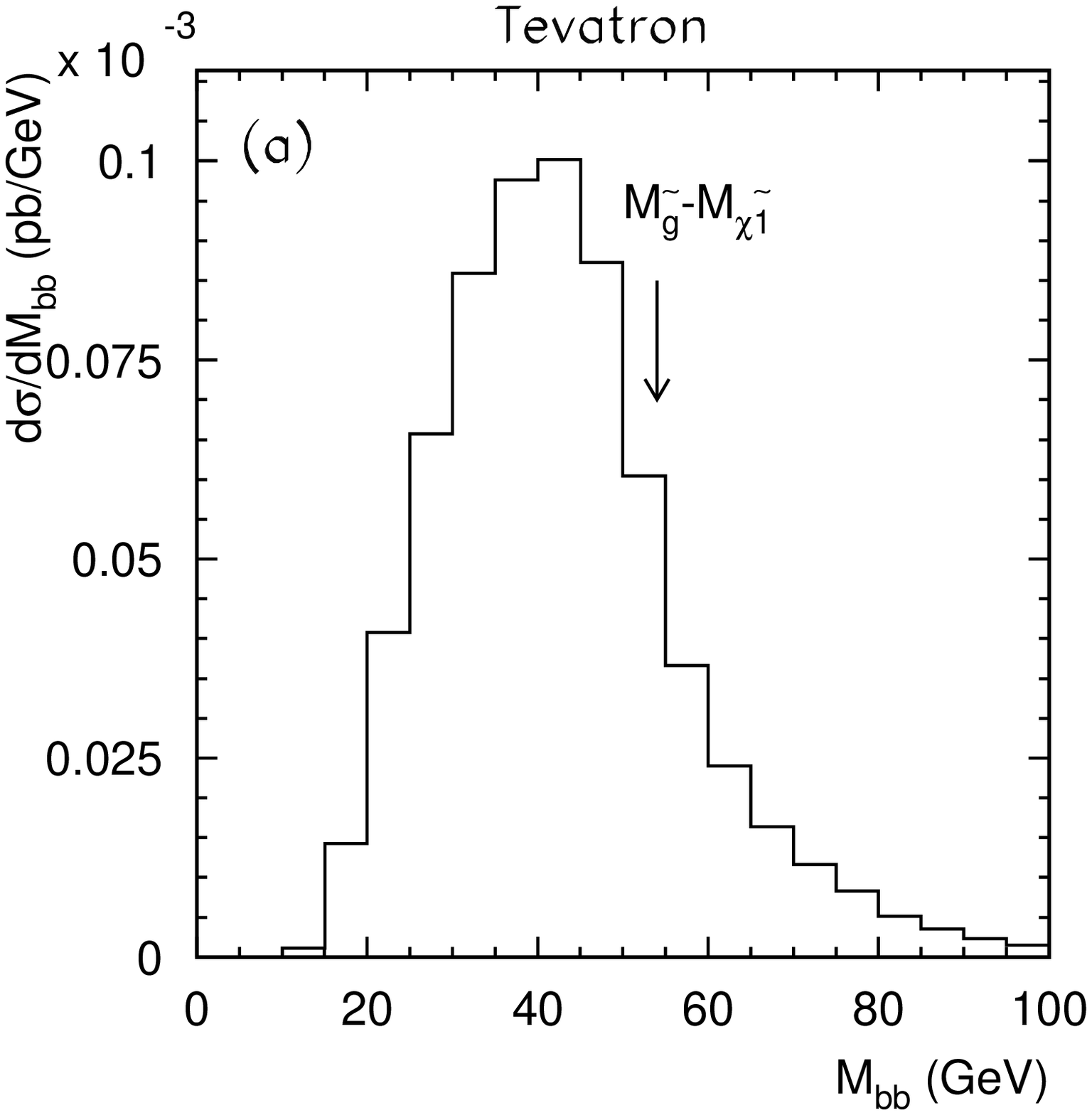}
\includegraphics[width=0.45\textwidth]{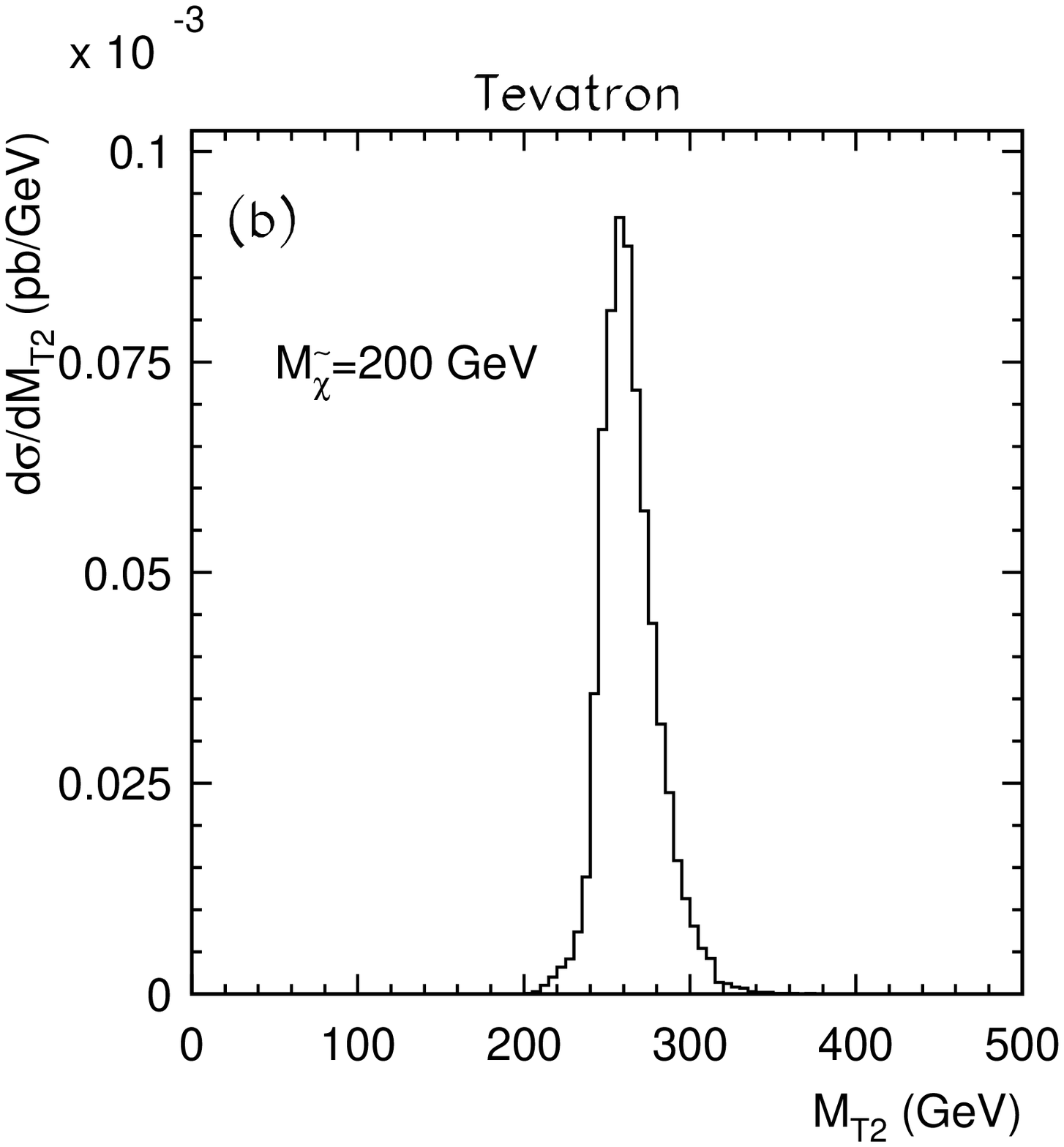}
\caption{(a) The distribution of di-$b$ jet invariant mass
$M_{b\bar{b}}$. (b) $M_{T2}$ reconstruction with 200 GeV trial LSP
mass. Both distributions are for Model B.} \label{dijet}
\end{figure}

In order to further identify the signal events and extract spectrum
information, we propose to reconstruct the events through the
invariant mass distribution of di-$b$ jets $M_{b\bar{b}}$ and
$M_{T2}$ method. There are 4 $b$-jets in the final states so we will
have three combinations of $b$-jet pair in event reconstruction.
Usually the two hardest $b$-jets come from different gluinos and the
$b$-jets pair coming from the same $\tilde{g}$ has smaller
separation angles. Therefore, following the reconstruction method in
Ref.~\cite{BaerYukawa}, we select the two hardest $b$-jets $b_1,b_2,$ and
let $b_3$ denote the third jet that minimizes $\Delta
\phi(b_2,b_3)$, such that the pair $b_2,b_3$ come from the same
$\tilde{g}$. The other $b$-jets pair from the second $\tilde{g}$
consists of two $b$-jets $b_1$ and $b_4$. For gluino three-body
decay, the dijet invariant mass $M_{b\bar{b}}$ must be less than the
mass difference between the gluino and the LSP
masses~\cite{Hinchliffe:1996iu} \beq M_{b\bar{b}} \leq \Delta M.
\eeq To illustrate this, we display the distribution of
$M_{b\bar{b}}=\max[M_{b_1b_4},M_{b_2b_3}]$ for benchmark model B in
Fig.~\ref{dijet}(a); it clearly shows the edge of $M_{b\bar{b}}$
around $\Delta M$. It is important to note that the final states
contain two invisible massive particles emanating separately from
two parent particles, and therefore it is usually hard to
reconstruct the signal events. However we still have sufficient
information to fully reconstruct the signal events. Since the signal
dominates according to our analyses above, the gluino mass can be
estimated from the production rate. Once $M_{\go}$ and the mass
difference $\Delta M$ are known, the LSP mass $M_{\tilde{\chi}_1^0}$
can be easily obtained. By substituting the LSP mass
$M_{\tilde{\chi}_1^0}$ into the reconstruction, we can plot the
variable $M_{T2}$ which is defined as~\cite{mt2,mt22} \beq
M_{T2}^2(\tilde{g}) = \min_{{\bf p}_T^{\chi^{(1)}}+{\bf
p}_T^{\chi^{(2)}}={\bf p}_T^{\rm
miss}}\{\max[M_T^{2(1)},M_T^{2(2)}]\}, \eeq where the transverse
masses $M_T^{(1)},M_T^{(2)}$ are constructed for two gluino decay
chains in terms of the relevant transverse invariant mass and
transverse momentum of $b\bar{b}$ system and as function of the
trial LSP mass. Figure~\ref{dijet}(b) shows the $M_{T2}$
reconstruction of the gluinos in Model B with trial LSP mass as
200~GeV. The consistency with our assumptions confirms the LSP mass
$M_{\tilde{\chi}_1^0}$ is the correct one and more importantly, it
can also be used as a check for the gluino-bino co-annihilation
condition as in Eq.~(\ref{co}), \beq M_{\tilde{\chi}_1^0} \simeq
M_{\go} / (1+ 20\%)\simeq 200~{\rm GeV}~. \eeq

\section{NLSP gluino Search at the LHC}

\begin{figure}[tb]
\centering 
\includegraphics[width=0.45\textwidth]{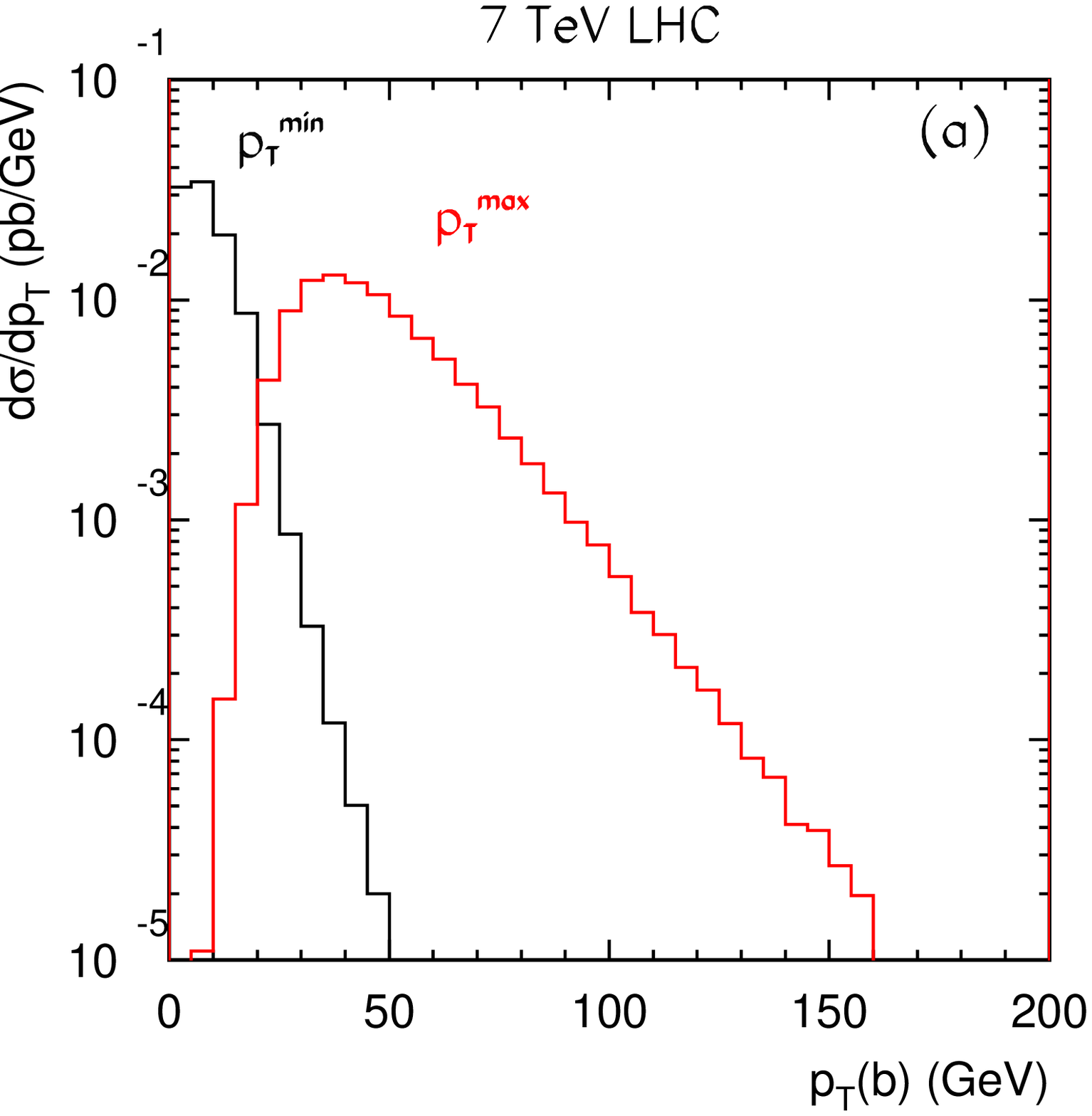}
\includegraphics[width=0.45\textwidth]{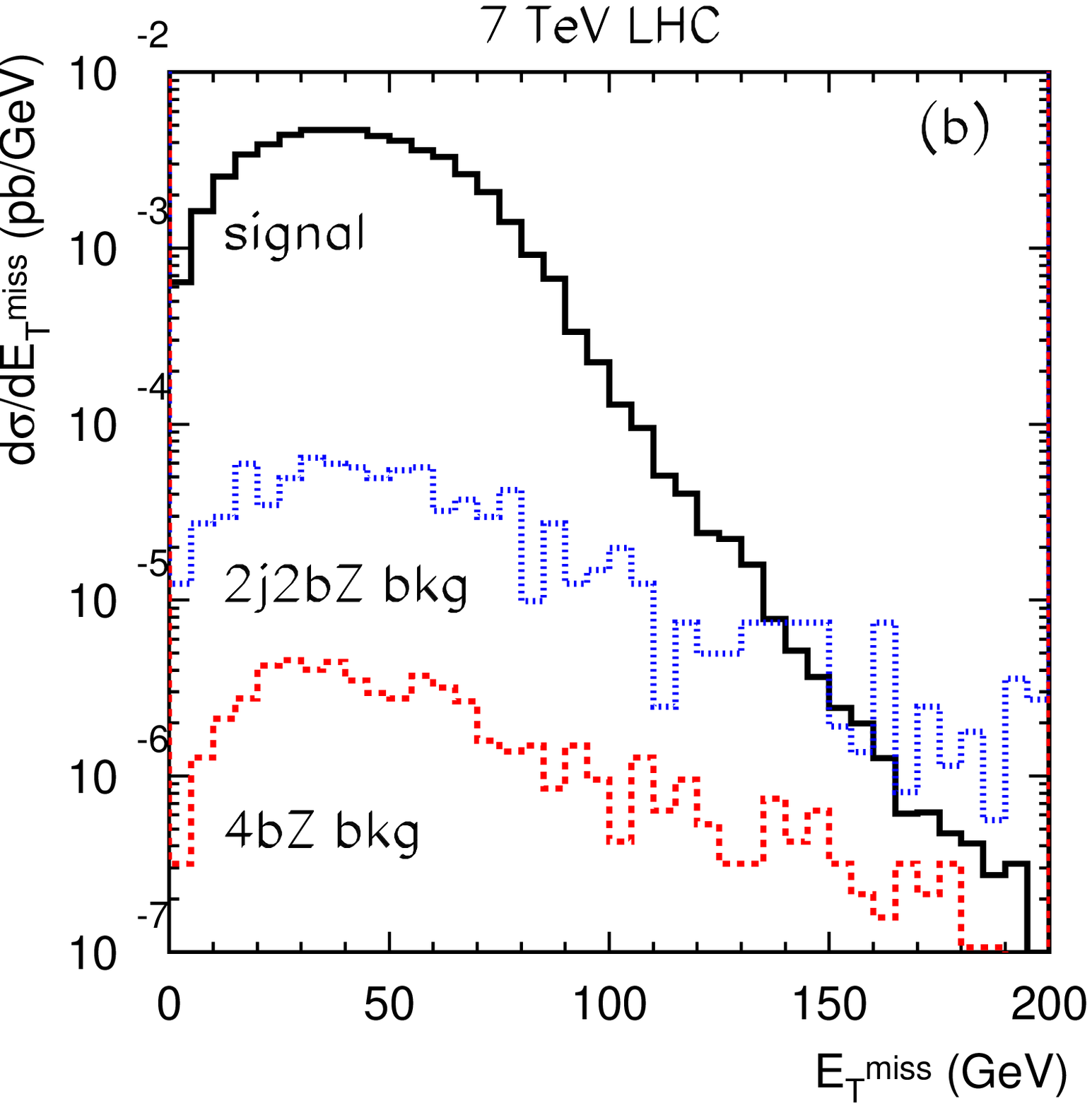}
\caption{(a) Minimal and maximal $p_{T}$ distribution of 4
$b$-jets in $pp\to \go\go\to
b\bar{b}b\bar{b}\tilde{\chi}^{0}_{1}\tilde{\chi}^{0}_{1}$ at 7 TeV
LHC. (b) Missing transverse energy ($\cancel{E}_{T}$) distribution
in signal events and in backgrounds $4bZ$ and $jjb\bar{b}Z$ at 7 TeV
LHC, with the branching fraction of $Z$ invisible decay
($Z\to\nu\bar{\nu}$) as 20\% and mis-$b$-tagging rate of light jet
as $1/30$.} \label{lhcdist}
\end{figure}

In this section we discuss the NLSP gluino search at the early LHC run of 7 TeV c.m. energy.
We employ a search strategy very similar to that for the Tevatron. In the LHC environment, the mis-$b$
tagging rate for light jets is about 1/30 for low $p_T$ range (15-50~GeV)~\cite{atlas}.  
However, since the gluino pair production at the LHC is dominated by $g g\to \go\go$, the total
cross section is about 30 times larger than at the Tevatron. We can then require
to tag all four $b$-jets and increase the missing transverse energy $\cancel{E}_T$ selection cut:
\begin{itemize}
\item 4 b-tagged jets with $p^{j}_{T} > 15$~GeV, $|\eta_{j}|< 2.0$, $\epsilon_b=50\%$~;
\item $\Delta R_{jj} > 0.4$~;
\item veto any event with lepton $p^{\ell}_{T}>10$~GeV~;
\item $\cancel{E}_{T}>40$~GeV~.
\end{itemize}
The smearing parameterization is given as~\cite{atlas}
\beq
\frac{\Delta E_{j}}{E_{j}}= \frac{50\%}{\sqrt{E_{j}/\rm GeV}}\oplus 3\%.
\eeq
We show the $p_{T}$ distribution of $b$-jet in the signal final states and $\cancel{E}_{T}$ distribution for signal and backgrounds in Fig.~\ref{lhcdist}(a) and (b) respectively.
One can see that the $p_T$ distribution of $b$-jets in the final states at the LHC is similar to that at the Tevatron. The missing transverse energy is smaller in comparison with the SM backgrounds. We summarize in Table~\ref{lhcresult} the results of signal and background
studies at the LHC for the two benchmark models. After all the selection cuts, the signal events far exceed the SM background.
By the first shut-down in winter 2010, the LHC should accumulate about 50 pb$^{-1}$ of data.
After the cuts, we expect negligible background events and 3 signal events for Model A, while Model B predicts about 7 events.
We expect our study to yield important clues about the underlying NLSP gluino scenario during the early stage of LHC operation.

At the LHC, another channel may also become interesting. With a light gluino exchange in the $t$-channel, the first generation squarks $\tilde{u}$, $\tilde{d}$ can be produced together with gluinos at significant rates via valence quark-gluon scattering. For Model B with $M_{\tilde{u}}\simeq M_{\tilde{d}}\sim 1.5$~TeV, the total
production cross section for
\beq
pp\to gu,gd \to \go \tilde{u},\go \tilde{d}
\eeq
is about 120~fb. With a NLSP gluino, $\tilde{q}\to \go q$ decay always
dominates and the gluinos
will be highly boosted. The signal from heavy squarks decay will consequently be two extremely hard jets with one of them being the collimated gluino. It is then difficult to tag
the $b$-jet in decay products of boosted gluino. However, the $b$-jets from the other gluino
decay can still be tagged. The heavy squark resonance
also provides a nice handle to suppress the SM backgrounds.
In addition, with the heavy resonance, this channel may enable the search
for $\go\to g\tilde{\chi}_1^0$ decay. In either case, the search will require
a more careful analyses of the jet substructure and we postpone this for future study.


\begin{table}[tb]
\begin{center}
\begin{tabular}[t]{|c|c|c|c|c|c|}
  \hline
  $\sigma(\rm fb) @$ 7 TeV LHC& Model A & Model B & $4b$ & $4bZ$ & $2j2bZ$\\
  \hline
  basic cuts and 4b tagging & 143 & 271 & $157\times 10^3$ & 0.55 & 4.2\\
  \hline
  $\cancel{E}_T>40$~GeV & 59 & 140 & $-$ & 0.4 & 3.3 \\
  \hline
\end{tabular}
\end{center}
\caption{Production cross section for signal $pp\to \tilde{g}\tilde{g}\to b\bar{b}b\bar{b}
\tilde{\chi}_1^0\tilde{\chi}_1^0$ and backgrounds $4b,4bZ,2j2bZ$ after basic cuts and missing energy cut at 7 TeV LHC for the two benchmark models.}
\label{lhcresult}
\end{table}

\section{Conclusion}
We have explored the collider phenomenology of gluino-bino co-annihilation scenarios
for both the Tevatron and 7 TeV LHC.
The NLSP gluino is only slightly more massive ($\sim 50$ GeV) than the bino LSP, so that
the conventional gluino searches $\tilde{\chi}_1^\pm \tilde{\chi}_1^\pm$+jets are not applicable. We propose to search
for gluino pairs through multi-b jets final states.
By using two benchmark points from a supersymmetric 4-2-2 model, in which NLSP gluino arises naturally, we explicitly show how
the search strategy works at hadron colliders. It is shown that with 10~fb$^{-1}$ integrated luminosity, one can reach over 4.5 $\sigma$ at Tevatron.
By the end of the first LHC run at 7~TeV with 50 pb$^{-1}$ of accumulated data, the predicted signal events
for both benchmark models are $\cal O$(10), with negligible SM background events.

\subsection*{Acknowledgments}
We would like to thank Ilia Gogoladze, Rizwan Khalid, Shabbar Rizvi,
Bruce Mellado and Tsutomu Yanagida for useful discussions. TL thanks
Tao Han for providing Fortran codes HANLIB for our calculations. MA,
TL and QS are supported by the DOE under grant No.
DE-FG02-91ER40626. KW is supported in part by the World Premier
International Research Center Initiative (WPI Initiative), MEXT,
Japan and the JSPS under grant JSPS Young Scientist (B) 22740143.

\renewcommand{\baselinestretch}{1.4}

\end{document}